\documentclass[11pt]{elsart}
\usepackage{graphics,color,epsfig,amsmath,amssymb}
\usepackage{feynmp}
\usepackage{subfigure}
\newcounter{bla}

\def\be{\begin{align}}
\def\ee{\end{align}}
\def\bea{\begin{align}}
\def\eea{\end{align}}
\def\nn{\nonumber}

\newcommand{\secdec}{{\textsc{SecDec}\,\,}}

\newcommand{\eps}{\epsilon}

\begin{document}

\begin{frontmatter}
\hfill{MPP-2013-48}\\ 

\title{Massive non-planar two-loop four-point integrals with SecDec 2.1}

\author[a]{Sophia Borowka},
\author[a]{Gudrun Heinrich}

\address[a]{Max-Planck-Institute for Physics, F\"ohringer Ring 6, 80805 M\"unchen, Germany}

\begin{abstract}
We present numerical results for massive non-planar two-loop box integrals
entering heavy quark pair production at NNLO,
some of which are not known analytically yet.
The results have been obtained with the program SecDec 2.1, 
based on sector decomposition and contour deformation,  
in combination with new types of transformations.
Among the new features of version 2.1 is also the possibility to evaluate contracted tensor 
integrals, with no limitation on the rank.

\begin{flushleft}
PACS: 12.38.Bx, 
02.60.Jh, 	
02.70.Wz 	
\end{flushleft}

\begin{keyword}
Perturbation theory, Feynman diagrams, multi-loop, numerical integration, top quark pair production
\end{keyword}

\end{abstract}

\end{frontmatter}

\newpage


\section{Introduction}
\label{sec:intro}

There are a few processes measured at the LHC where the need for corrections 
beyond next-to-leading order (NLO) is free of doubt. 
One of them is top quark pair production, 
and the completion of a full NNLO prediction
is  well under way~\cite{Czakon:2012pz,Czakon:2012zr,Baernreuther:2012ws,Bonciani:2010mn,Bonciani:2009nb,Bonciani:2008az,AvMACAT,vonManteuffel:2012je,Abelof:2012he,Abelof:2012rv,Abelof:2011ap,Korner:2008bn,Aliev:2010zk}.
Soft gluon and Coulomb effects also have been taken into account beyond the next-to-leading logarithmic 
accuracy
and have been combined with fixed order results to come up with predictions as precise as 
possible~\cite{Czakon:2013vfa,Moch:2012mk,Cacciari:2011hy,Czakon:2011xx,Beneke:2011mq,Beneke:2012wb,Ahrens:2011px,Kidonakis:2010dk}.\\
Among the key ingredients of the full NNLO calculation are the master integrals entering the two-loop 
virtual corrections.
While in \cite{Czakon:2008zk,Czakon:2007ej,Czakon:2007wk} the latter enter in a semi-numerical form, 
a fully analytical representation for a subset of the needed master integrals has been 
presented 
in~\cite{Bonciani:2010mn,Bonciani:2009nb,Bonciani:2008az}, where the fermionic and leading colour
contributions to the $q\bar{q}$ channel and the leading colour contributions
to the $gg$ channel are calculated.
An analytical representation for the non-planar seven-propagator integral occurring 
in the light fermionic correction to the $gg$ channel 
also has been achieved\,\cite{AvMACAT,vonManteuffel:2012je}.
However, explicit results for some of the most complicated non-planar master integrals are still missing.

Here we present numerical results for two  non-planar seven-propagator topologies, 
one entering  the light fermionic correction to the $gg$ channel,  
where analytical results exist\,\cite{AvMACAT,vonManteuffel:2012je}, the other one 
entering  the heavy fermionic correction to the $gg$ channel, where no analytical result is 
available yet. Apart from scalar master integrals, we also give results for an irreducible 
tensor integral of rank two for the diagram entering  the heavy fermionic correction to the $gg$ channel.
In order demonstrate the 
applicability of the tensor option to various types of integrals, we also calculate some 
two-loop two-point functions involving several different mass scales and show that the timings 
for the tensor integrals are not much larger than the ones for the scalar integrals. 
This means that a numerical approach in certain cases can help to alleviate or even avoid the 
procedure of amplitude reduction  to master integrals.

The results have been obtained with the program \secdec2.1~\cite{Carter:2010hi,Borowka:2012yc,Borowka:2012rt}, 
based on sector decomposition to disentangle the singularity structure, followed by 
numerical contour integration. Compared to version 2.0 of \secdec, version 2.1 contains a number of new
features, which are also presented in this article. Among them is the possibility to evaluate 
tensor integrals with (in principle) no limitation on the rank. 
Another new feature is the option to apply the sector decomposition algorithm and subsequent contour 
deformation on user-defined functions which do not necessarily have the form of standard loop integrals. 
In fact, to achieve a convenient representation for one of the seven-propagator integrals, 
some analytical manipulations have been done before starting the algorithm, this way 
reducing the number of produced subsectors, leading to improved numerical behaviour. 
 
\medskip

The structure of this article is as follows. 
In Section \ref{sec:ggtt} we will derive the expression serving as a starting point 
for the evaluation of the massive non-planar two-loop box diagram mentioned above,
and describe novel types of transformations which can be used to reduce the number of 
sector decompositions.
The new 
features of the program \secdec2.1  are discussed in Section \ref{sec:program}. 
Numerical results are presented in Section \ref{sec:results}, where we also 
give results for some rank three tensor integrals for massive two-point functions.
An appendix contains a manual-style description of the new features of the program.
Detailed documentation of the program also comes with the code, 
which is available at {\tt http://secdec.hepforge.org}.

\section{Analytic preparation of the master diagrams}
\label{sec:ggtt}
The structure and usage of \secdec and the procedures it uses are described in detail in 
Refs. \cite{Carter:2010hi,Borowka:2012yc,Binoth:2000ps,Heinrich:2008si}. 
The main purpose of this section is to explore the possibilities arising from 
a mixed approach, where simple analytic manipulations to the integral 
before feeding it to the sector decomposition algorithm can lead to a large gain 
in efficiency for the subsequent numerical evaluation.

\subsection{Non-planar seven propagator integrals with two massive on-shell legs}
\begin{figure}[h]
\subfigure[ggtt1]{\raisebox{8pt}{\includegraphics[width=0.45\textwidth]{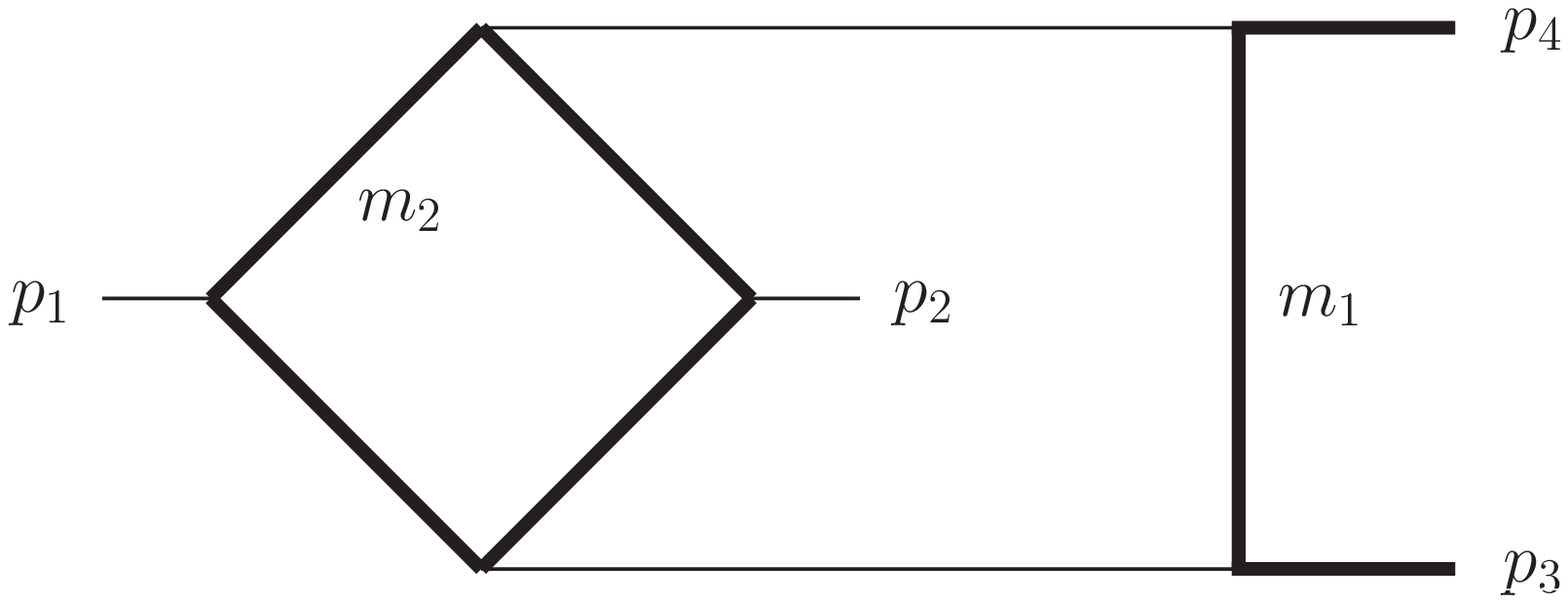}} }\hfill
\subfigure[ggtt2]{\includegraphics[width=0.4\textwidth]{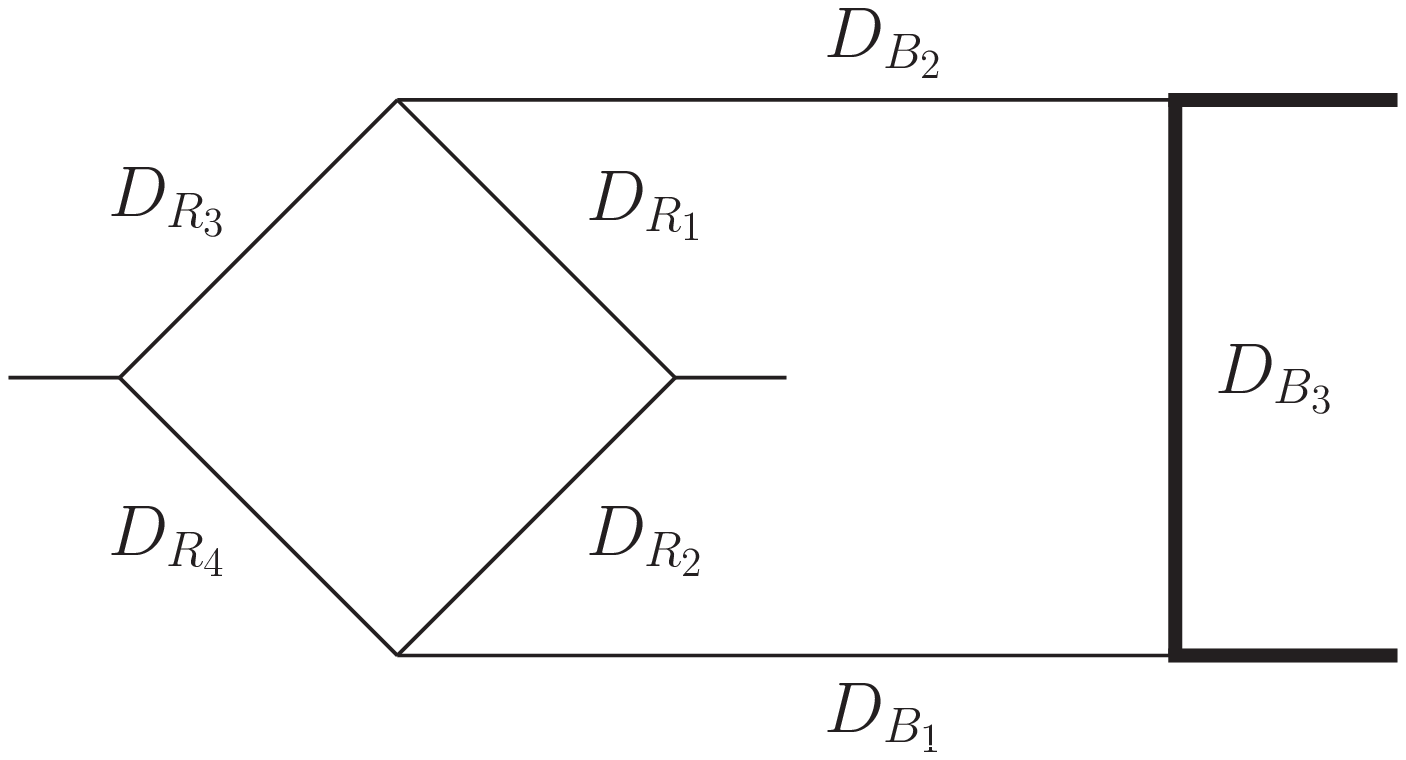} }
	\caption{Massive non-planar two-loop box diagrams entering the heavy (a) and 
	light (b) fermionic correction to the $gg$ channel; the thick lines denote massive particles.} 
	\label{fig:ggtt2} 
\end{figure}

The diagrams shown in Fig. \ref{fig:ggtt2} are master topologies occurring in the two-loop corrections 
to $t\bar{t}$ production in the $gg$ channel. While results for 
the diagram corresponding to massless fermionic corrections 
in a sub-loop -- which we call {\it ggtt2} --  are available in terms of ${\cal O}(800)$ generalized 
polylogarithms~\cite{AvMACAT}, analytic results for the integral {\it ggtt1}, 
containing a sub-diagram with a massive loop, are not available.
Numerically however, the evaluation of {\it ggtt1} is easier than the one of {\it ggtt2}, 
due to its less complicated infrared singularity structure. 
While the leading poles of {\it ggtt2} are of order $1/\eps^{4}$, 
and intermediate expressions during sector decomposition 
show (spurious) pole structures where the degree of divergence is higher than logarithmic,
the integral {\it ggtt1} has only finite contributions.
Therefore we  evaluate {\it ggtt1} with \secdec\,2.1 fully automatically, while for {\it ggtt2}
it is advantageous to make some analytical manipulations beforehand, 
reducing both, the number of Feynman parameters to be integrated over numerically 
and the degree of divergence.

The expression for the scalar integral {\it ggtt2}  in momentum space is given by
\begin{align}
\label{eq:gnp}
 \mathcal{G}_{ggtt2}= \left( \frac{1}{\mathrm{i} \pi^{\frac{\mathrm{D}}{2}}}\right)^2 
 \int \frac{\mathrm{d^D}k_1\,\, \mathrm{d^D}k_2}{D_{R_1}D_{R_2}D_{R_3}D_{R_4}D_{B_1}D_{B_2}D_{B_3}}  
\end{align}
where $\mathrm{D}=4-2 \epsilon$.
The  Feynman propagators $D_{R_i}$ corresponding to the ``rhombus" sub-loop in Fig. \ref{fig:ggtt2} 
are given by
\begin{align}
 D_{R_1} &= (k_1-k_2)^2+\mathrm{i}\delta \text{ ,}\hspace{15pt}  D_{R_2}= (k_1-k_2+p_2)^2+
 \mathrm{i}\delta \text{ ,}\nn\\
 D_{R_3} &= (k_2+p_4)^2+\mathrm{i}\delta \text{ ,}\hspace{15pt} D_{R_4}= (k_2+p_1+p_4)^2+
 \mathrm{i}\delta \text{ ,}
\end{align}
where the $p_i$ are the external momenta with $p_3^2=p_4^2=m^2$ and $p_1^2=p_2^2=0$, 
and $k_1$, $k_2$ are the loop momenta. 
Integrating out the loop momentum $k_2$ first, we are left with an expression containing 
$k_1$ and external momenta only, to be combined with the propagators 
\begin{align}
 D_{B_1} = (k_1-p_3)^2+\mathrm{i}\delta \text{ ,}\hspace{10pt}  D_{B_2}= (k_1+p_4)^2+
 \mathrm{i}\delta \text{ ,}\hspace{10pt} D_{B_3} = k_1^2-m^2+\mathrm{i}\delta \;.
 \label{eq:Dbs}
\end{align}

The introduction of Feynman parameters for the one-loop subgraph $\mathcal{I}_{R}$ 
containing only the loop momentum $k_2$ leads to
\begin{align}
 \mathcal{I}_{R} = \frac{1}{\mathrm{i} \pi^{\mathrm{D}/2}} \int \frac{ \mathrm{d^D}k_2 }
 { D_{R_1}D_{R_2}D_{R_3}D_{R_4}}
 = \Gamma(2+\epsilon) \int \prod_{i=1}^4\mathrm{d}x_i \delta(1-\sum_{j=1}^4 x_j) 
 \mathcal{F}(\vec{x},k_1)^{-2-\epsilon} \text{ ,}\label{loop1}
\end{align}
with
\begin{align}
\!\! -\mathcal{F}(\vec{x},k_1)= D_{B_1} x_1x_2 + (k_1+p_1+p_4)^2x_1x_3 + 
(k_1+p_2+p_4)^2 x_2x_4 +  D_{B_2} x_3x_4 \text{ .}\nn
\end{align}
We eliminate the $\delta$-function in eq.\,(\ref{loop1}) with the substitution
\begin{align}
\label{eq:parametrisation}
x_1 = t_2 (1- t_3) \text{ ,}\hspace{19pt} x_2 = t_1 t_3 \text{ ,}\hspace{19pt} x_3 = (1- t_1) t_3 \text{ ,}
\hspace{19pt}
\end{align}
to achieve a factorisation of the parameter $t_3$ which then is integrated out analytically, leading to
\begin{align}
 \mathcal{I}_{R} &= -\frac{2}{\epsilon}\frac{\Gamma(2+\epsilon)\Gamma^2(1-\epsilon)}{\Gamma(1-2\epsilon)} 
 \int_0^1 \mathrm{d}t_1\int_0^1\mathrm{d}t_2 \, \, \mathcal{\tilde F}(\vec{t},k_1)^{-2-\epsilon}\nn
\end{align}
with
\begin{align}
-\mathcal{\tilde F}(\vec{t},k_1)= D_{B_1} t_1t_2 + (k_1+p_1+p_4)^2 t_1\bar{t}_2+ 
(k_1+p_2+p_4)^2 \bar{t}_1t_2 +  D_{B_2} \bar{t}_1\bar{t}_2 \text{ ,}
\label{eq:ftilda}
\end{align}
and where we used the shorthand notation $\bar{t}_i=1-t_i$. %
Now we combine the expression for the 1-loop rhombus $\mathcal{I}_{R}$ 
with the remaining $k_1$-dependent propagators, 
treating the expression of eq.~(\ref{eq:ftilda}) as a fourth propagator with power 
$2+\epsilon$, to obtain, after integrating out $k_1$, 
\begin{align}
\nonumber \mathcal{G}_{NP}=&\frac{2}{\epsilon}\frac{\Gamma(3+2\epsilon)\Gamma^2(1-\epsilon)}{\Gamma(1-2\epsilon)} \int_0^1 \mathrm{d}t_1\int_0^1\mathrm{d}t_2 \,\, \times\label{GNP} \\
 &\prod_{i=1}^4\int_0^1 \mathrm{d}z_i \,z_4^{1+\epsilon} \,\delta(1-\sum_{j=1}^4 z_j) \,\,\mathcal{F_{NP}}(\vec{z},t_1,t_2)^{-3-2\epsilon} \,\,\mathcal{U_{NP}}(\vec{z})^{1+3\epsilon} \text{ ,}
\end{align}
where 
\begin{align}
\mathcal{U_{NP}}(\vec{z})&= \sum_{j=1}^4 z_j \hspace{19pt}\text{ and} \nn\\
\label{eq:Fnp}
\mathcal{F_{NP}}(\vec{z},t_i)&= -s_{12} z_2 z_3 - T z_1 z_4 - S_1 z_2 z_4 -S_2 z_3 z_4 + m^2 z_1 (z_1
+z_4 Q)  \text{ ,}
\end{align}
with
\begin{align}
\nonumber T &= s_{13} \bar{t}_1 t_2 +s_{23} t_1 \bar{t}_2 \text{ ,}\hspace{19pt}  S_1= s_{12} t_1 t_2 \text{ ,}\hspace{19pt} S_2=s_{12} \bar{t}_1 \bar{t}_2 \\
\label{eq:STUdefs}
 Q &=t_1\bar{t}_2+\bar{t}_1t_2\text{ ,}\hspace{19pt} s_{ij} = (p_i+p_j)^2\text{ .}
\end{align}
Now we eliminate the $\delta$-function by performing a 
primary sector decomposition~\cite{Binoth:2000ps} in $z_1,\dots,z_4$ to obtain
\begin{align}
\mathcal{G}_{NP}=&\frac{2}{\epsilon}\frac{\Gamma(3+2\epsilon)\Gamma^2(1-\epsilon)}{\Gamma(1-2\epsilon)} \int_0^1 \mathrm{d}t_1\int_0^1\mathrm{d}t_2 \sum_{i=1}^4 G_{NP}^i \text{ ,}
\end{align}
with
\begin{align}
G_{NP}^1&=\int_0^1 \mathrm{d}z_2\, \mathrm{d}z_3 \,\mathrm{d}z_4 \,\,\,z_4^{1+\epsilon}\,\,\,(1+z_2+z_3+z_4)^{1+3 \epsilon}\,\,\,\mathcal{F}^1(\vec{z},t_i)^{-3-2\epsilon} \nn\\
\mathcal{F}^1(\vec{z},t_i)&= -s_{12} z_2 z_3 - T z_4 - S_1 z_2 z_4 -S_2 z_3 z_4 + m^2 (1 +z_4 Q)  \text{ ,}\nn\\
G_{NP}^2&=\int_0^1 \mathrm{d}z_1 \,\mathrm{d}z_3 \,\mathrm{d}z_4 \,\,\,z_4^{1+\epsilon}\,\,\,(1+z_1+z_3+z_4)^{1+3 \epsilon}\,\,\,\mathcal{F}^2(\vec{z},t_i)^{-3-2\epsilon}\nn \\
\mathcal{F}^2(\vec{z},t_i)&= -s_{12} z_3 - T z_1 z_4 - S_1 z_4 -S_2 z_3 z_4 + m^2 z_1 (z_1 +z_4 Q)  \text{ ,}\label{eq:Fsec2}\\
G_{NP}^3&=\int_0^1 \mathrm{d}z_1\, \mathrm{d}z_2 \,\mathrm{d}z_4 \,\,\,z_4^{1+\epsilon}\,\,\,(1+z_1+z_2+z_4)^{1+3 \epsilon}\,\,\,\mathcal{F}^3(\vec{z},t_i)^{-3-2\epsilon} \nn\\
\mathcal{F}^3(\vec{z},t_i)&= -s_{12} z_2 - T z_1 z_4 - S_1 z_2 z_4 -S_2 z_4 + m^2 z_1 (z_1 +z_4 Q)  \text{ ,}\nn\\
G_{NP}^4&=\int_0^1 \mathrm{d}z_1\, \mathrm{d}z_2 \,\mathrm{d}z_3 \,\,\,(1+z_1+z_2+z_3)^{1+3 \epsilon} \,\,\,\mathcal{F}^4(\vec{z},t_i)^{-3-2\epsilon} \nn\\
\mathcal{F}^4(\vec{z},t_i)&= -s_{12} z_2 z_3 - T z_1 - S_1 z_2 -S_2 z_3 + m^2 z_1 (z_1 + Q)  \text{ .}
\end{align}
We observe that $\mathcal{F}^1(\vec{z},t_i)$ is of the form $m^2+{\mathrm{func}}(z_i,t_i)$, 
so does not need any further decomposition, and primary sector 3 can be remapped to primary sector 2 
by exchanging $z_2 \leftrightarrow z_3$ and $S_1 \leftrightarrow S_2$. 
Hence, we are left with the treatment of primary sectors 2 and 4 only. 
\\
The integrals $G_{NP}^{2,3,4}$ can have singularities both at zero and one in $t_1$ and $t_2$. 
With the sector decomposition algorithm, only singularities at zero are factorized automatically. 
Consequently, we remap the singularities located at the upper integration limit 
to the origin of parameter space by splitting the integration region at $\tfrac{1}{2}$
and transforming the integration variables to remap the integration domain to the unit cube\,\cite{Heinrich:2008si}.
This procedure results in 12 integrals,   
some of which already being finite, such that no subsequent sector decomposition is required.
Some of the integrals however lead to singularities of the type $\int_0^1dx\,x^{-2-\eps}$ after sector decomposition, 
which we call {\it linear } singularities. These singularities are spurious and can be subtracted 
by expanding the Taylor series in the subtraction procedure up to the second term~\cite{Binoth:2000ps,Binoth:2003ak}.
However, this procedure can lead to large cancellations between subtraction terms and therefore to numerical 
instabilities. Hence it is advisable to try  avoiding this type of singularity from scratch.
In the following section we describe a method which can help to do so.

\subsection{Removal of double linear divergences via backwards transformation}
\label{subsec:backsecdec}
The aim of the procedure described in this section is to achieve a transformation of 
potential linear divergences into logarithmic 
divergences as far a possible. 
A different procedure towards this goal, based on integration-by-parts identities, 
has been described in \cite{Carter:2010hi}. 
The latter method however can increase the number of functions 
to be integrated substantially, 
while the method described below in general reduces the number of further iterations
and therefore the number of produced functions.
Yet another method to reduce the number of functions produced during factorization 
has been suggested in Ref.\,\cite{Anastasiou:2010pw}, but we do not use it here as 
it can introduce singularities at the upper integration limit which subsequently 
have to be remapped again.

To explain the type of transformation advocated here,  
we use a  function of the structure of $\mathcal{F}^2$ in eq.~(\ref{eq:Fsec2}) as an example, 
renaming $t_2\to z_2, t_1\to z_5$.
Concerning the Feynman parameters, we can identify the following structure in 
eq.~(\ref{eq:Fsec2}) (in our concrete example $N=5, z_j=z_4, z_k=z_1$)
\begin{align}
 I= \prod_{i=1}^N\left\{\int_0^1  {\rm d}z_i \right\}\,
 \left[ z_j\left( P(\vec{z}_{jk}) + z_k Q(\vec{z}_{jk}) \right) + R(\vec{z}_{jk}) \right]^{- \alpha} \text{ ,}
\label{eq:secback}
\end{align}
where  $\alpha$ is assumed to be positive, and where $P,Q$ and $R$ are 
polynomials of arbitrary degree of the Feynman parameters 
$\vec{z}_{jk}=(z_1,\dots,\hat{z}_j,\dots,\hat{z}_k,\dots,z_N)$ and kinematic invariants. 
The hat denotes Feynman parameters
which do {\it not} occur in the corresponding function.\\
In eq.~(\ref{eq:secback}), all terms multiplied by the Feynman parameter $z_k$ are 
also multiplied by the Feynman parameter $z_j$. Hence, the sector decomposition method can be applied ``backwards".
To explain this in more detail, consider the following function:
\begin{align}
J=& \prod_{i=1}^N\left\{\int_0^1  {\rm d}z_i \right\}\,\frac{1}{z_j}
\left[ z_j P(\vec{z}_{jk}) + z_k Q(\vec{z}_{jk}) + R(\vec{z}_{jk})\right]^{- \alpha} [\underbrace{\Theta(z_k-z_j)}_{(1)}+\underbrace{\Theta(z_j-z_k)}_{(2)}]\label{eq:wayback}
\end{align}
Now we substitute $z_j=z_k\,t_j$ in sector (1) and  $z_k=z_j\,t_k$ in sector (2), to obtain, 
after renaming again $t_i$ into $z_i$
\begin{align}
J &= \prod_{i=1}^N\left\{\int_0^1  {\rm d}z_i \right\}\frac{1}{z_j}
\left[ z_j P(\vec{z}_{jk}) + z_k Q(\vec{z}_{jk}) + R(\vec{z}_{jk})\right]^{- \alpha} \label{eq:backwards0}\\
&=\prod_{i=1}^N\left\{\int_0^1  {\rm d}z_i \right\}\frac{1}{z_j}
\left[ z_k (z_j P(\vec{z}_{jk}) + Q(\vec{z}_{jk}) ) + R(\vec{z}_{jk})\right]^{- \alpha} \label{eq:backwards1}\\
&\quad+\prod_{i=1}^N\left\{\int_0^1  {\rm d}z_i \right\}
\left[ z_j ( P(\vec{z}_{jk}) + z_k Q(\vec{z}_{jk}) ) + R(\vec{z}_{jk})\right]^{- \alpha} \;.\label{eq:backwards2}
\end{align}
We observe that the  term in eq.~(\ref{eq:backwards2}) is the same as eq.~(\ref{eq:secback}). 
Therefore we can replace it by the expressions (\ref{eq:backwards0}) minus (\ref{eq:backwards1}). 
The effect is twofold: The degree of the polynomial in $z_jz_k$  is reduced  in eq.~(\ref{eq:backwards0}),  
and  in eq.~(\ref{eq:backwards1})  the degree of divergence
in $z_j$ is reduced if $\alpha > 1$. 
\vspace{14pt}\\
After all transformations of this type we arrive at a total of 15 functions partly needing an iterated sector decomposition. 
Together with the introduction of the new feature of {\it user-defined functions} in \secdec 2.1,
which we will describe in the following section, 
we are now able to compute the {\it ggtt2} diagram in a reasonable amount of time.

\section{Structure and new features  of \secdec version 2.1}
\label{sec:program}
\subsection{Structure}
\label{subsec:structure}
\begin{figure}[htb]
	\begin{center}
	      \includegraphics[width=1.0\textwidth]{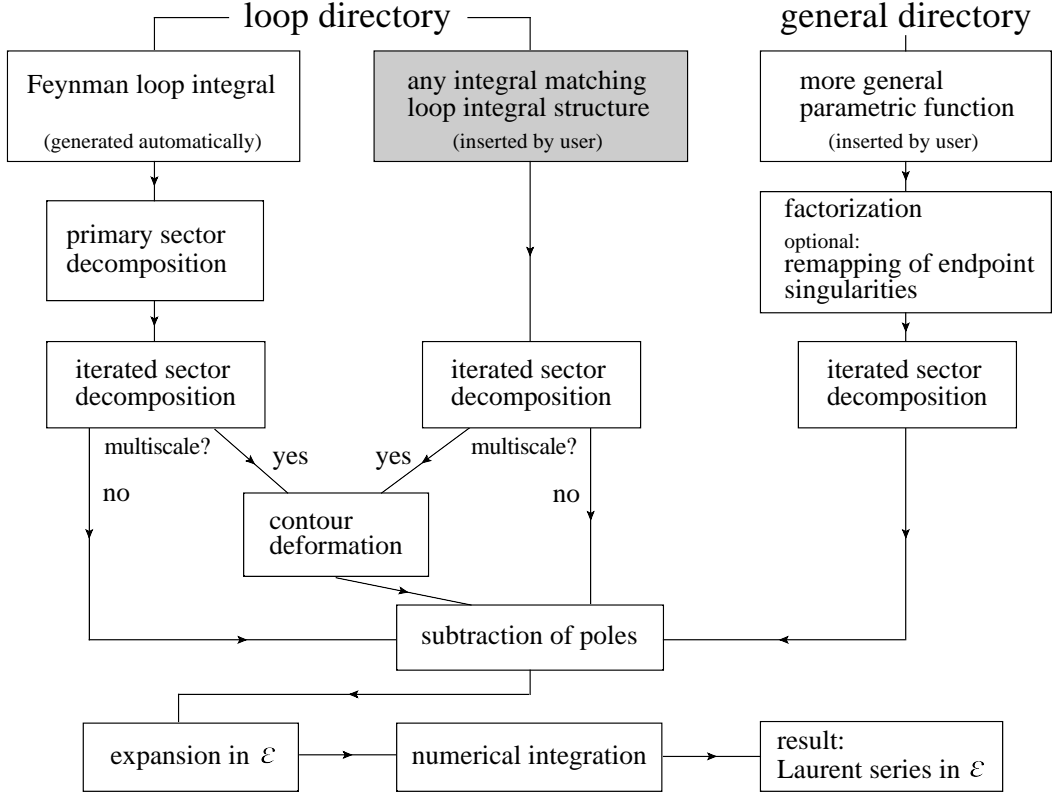}
	\end{center}
	\caption{Flowchart showing the main steps the program performs to produce the 
	numerical result as a Laurent series in $\epsilon$.}
	\label{fig:structure} 
\end{figure}

The workflow of the program is shown in Fig.~\ref{fig:structure}. The directory structure of \secdec
splits into two main branches: {\tt loop} and {\tt general}. 
In the {\tt loop} part, multi-scale loop integrals can be evaluated without restricting the kinematics to the Euclidean region. Integrable singularities are dealt with by deforming the 
integration contour into the complex plane.
In the {\tt general} part, the poles of more general parametric functions can be factorized, 
and subsequently the functions can be evaluated numerically, provided they contain  
only end-point singularities. Contour deformation is not available in the subdirectory {\tt general}.
For more information on the structure of the various subdirectories we 
refer to Ref.~\cite{Borowka:2012yc}.

\subsection{New features}
\label{subsec:news}
The main new features of \secdec version 2.1 are the {\it -u} option in the {\tt loop} directory
to decompose ``user defined" functions rather than 
standard multi-loop integrals,  
an extended tensor integral option, and improvements in the error treatment.

\subsection*{Evaluation of user-defined functions with arbitrary kinematics}

If the user would like to calculate a ``standard" loop integral, it is sufficient 
to specify the propagators, and the program will construct the integrand in terms of 
Feynman parameters automatically.
However, in section \ref{sec:ggtt} we have seen that an analytical step can be helpful when dealing with 
complicated integrals. 
Integrating out one Feynman parameter analy\-ti\-cally reduces the number of integration variables for the subsequent
Monte Carlo integration and therefore in general improves numerical efficiency. 
This implies that the constraint $\delta(1-\sum_i x_i)$ has been used already to achieve a 
convenient parametrisation, and therefore no primary sector decomposition 
to eliminate the $\delta$-constraint is needed anymore.
In such a case, the user can skip the primary sector decomposition step and 
insert the functions to be factorized directly into the Mathematica input file, 
using his favourite parametrisation.
More generally speaking, the purpose of this option is to be more flexible with regards to the 
functions to be integrated, such that expressions for loop integrals which are not in the ``standard form" -- 
for example due to 
analytic manipulations which have been performed already on the integral -- can be dealt with as well.
This includes the possibility to perform a deformation of the integration contour into the complex plane, 
taking the user-defined functions as a starting point.
Oriented at the functions $\mathcal{F}$ and $\mathcal{U}$ for 
the ``standard" loop case (see e.g. eq.~(\ref{GNP})), 
the user-defined functions can encompass the product of two arbitrary 
polynomial functions with different exponents and an additional numerator.
Details about the usage of this option are given in appendix \ref{appendix:userdefined}. 

The tensor integral option and the syntax for the definition of the numerator in terms of 
contracted loop momenta 
are described in detail in appendix \ref{appendix:tensor}.

\subsection*{Error estimates} 
When dealing with complicated integrands it can happen that the error given by the numerical integration program -- which is based on 
the number of sampling points only -- underestimates the true error.
The numerical integrators contained in the {\scshape Cuba} library~\cite{Hahn:2004fe,Agrawal:2011tm} 
give an estimate of the correctness of the
 stated error to a given result.
The new \secdec version 2.1 collects the maximal error probability for each computed order in the dimensional regulator $\epsilon$ and
 writes it to the result files {\tt *.res}. 
In the generic case, the information on the reliability of 
the stated error is given as a probability with values between 0 and 1. 
If the integrator returns a value larger than one, 
the integration has not come to  successful completion, 
and a warning is written to the result files. 
This feature helps the user to judge how reliable the numerical results, respectively the given errors, are.
\subsection{Installation and usage}
\label{subsec:install}
The program can be downloaded from {\tt http://secdec.hepforge.org}.\\
Unpacking the tar archive via {\it  tar xzvf SecDec.tar.gz} 
will create a directory called {\tt SecDec} 
with the subdirectories as described above. 
Installation is done by changing to the {\tt SecDec} directory and running {\it ./install}.

Prerequisites are Mathematica, version 6 or above, Perl (installed by default on 
most Unix/Linux systems), a C++ compiler, and a Fortran compiler if the Fortran option is used.

Details about the usage can be found in the appendix, where manual-type descriptions 
of the new features are  given, and also in Ref.~\cite{Borowka:2012yc} 
and the documentation coming with the program.

\section{Results}
\label{sec:results}
Now we present numerical results for two non-planar seven propagator master 
topologies occurring in the two-loop corrections to the process $gg\to t\bar{t}$, 
shown in Fig.~\ref{fig:ggtt2}.
In addition to the scalar integrals, we also compute an irreducible tensor integral.
In order to demonstrate the applicability of the tensor option to various contexts, 
we also give results for some rank three tensor two-point functions involving several mass scales.

\subsection{The {\it ggtt1} diagram}

\begin{figure}[htb]
\subfigure[scalar integral]{\includegraphics[width=7.cm]{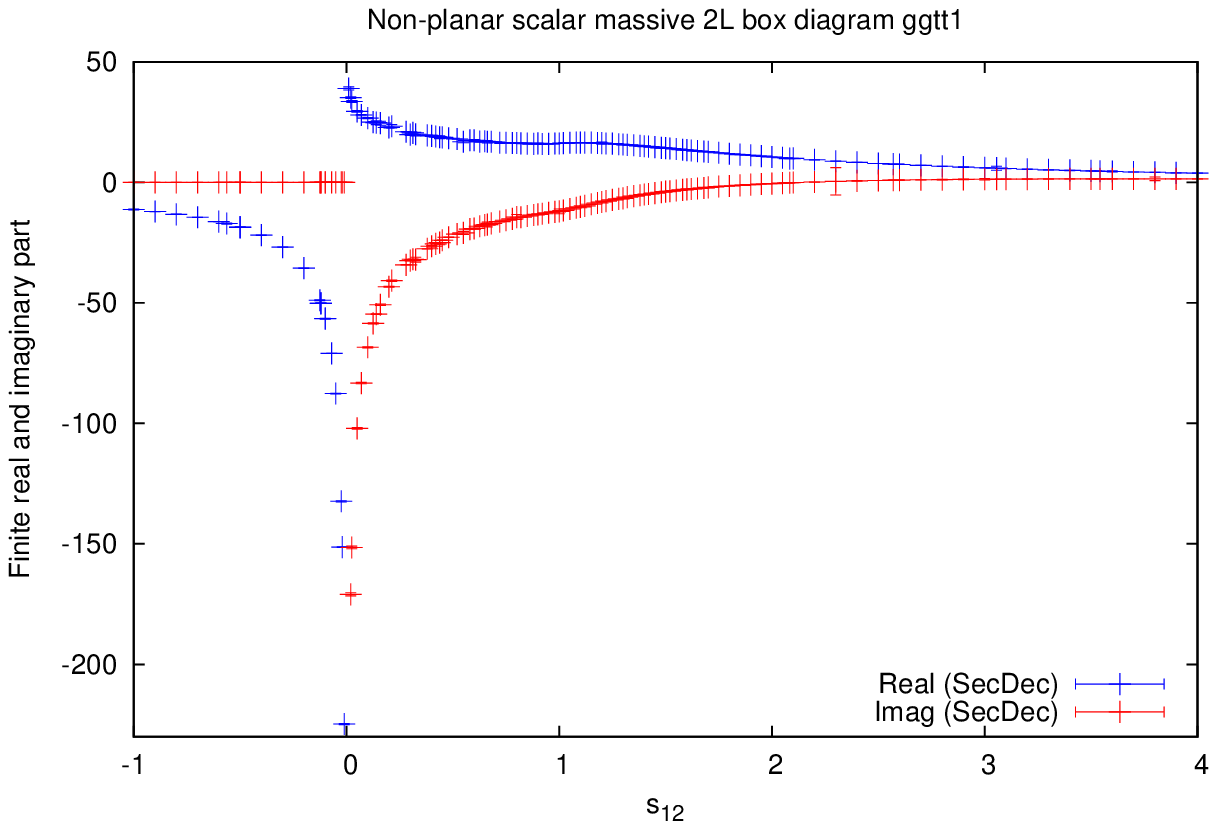} }\hfill
\subfigure[rank 2 tensor integral]{\includegraphics[width=7.cm]{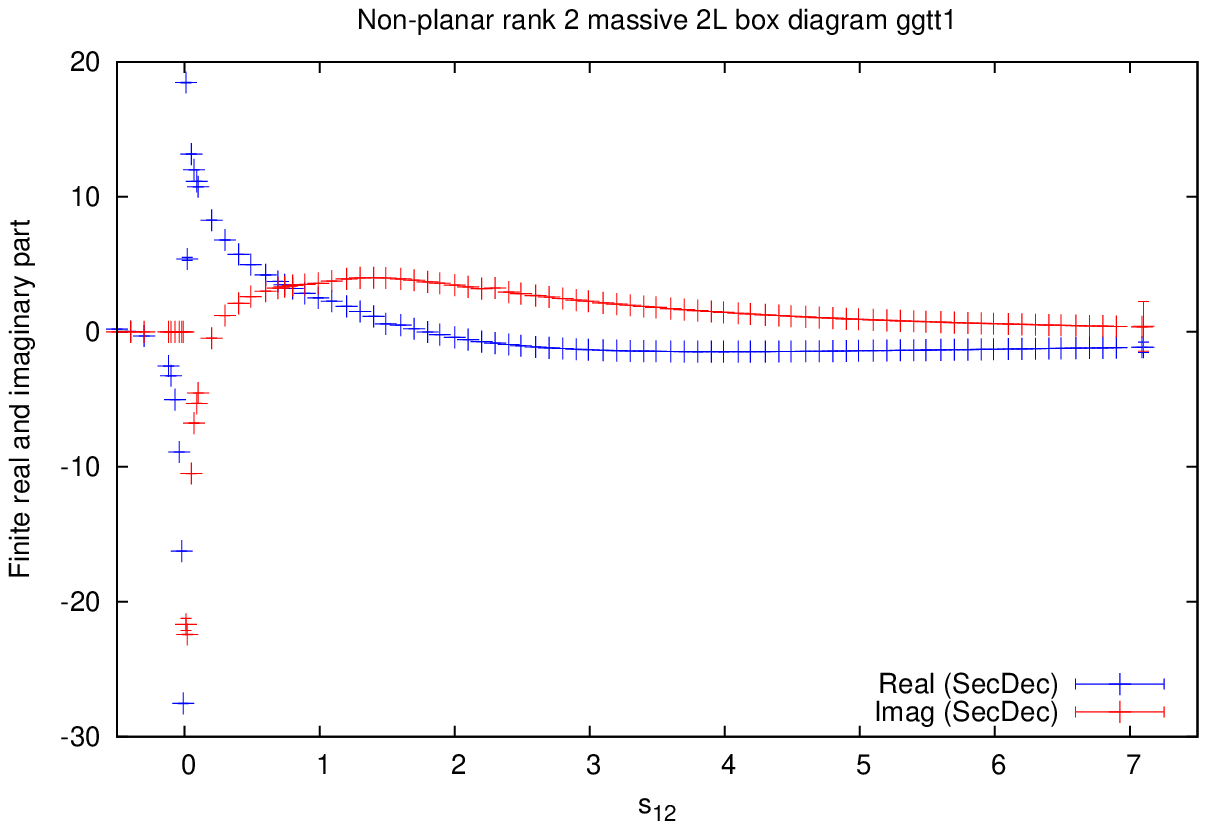} }
\caption{Results for the scalar integral  {\it ggtt1} shown in Fig.~\ref{fig:ggtt2}(a), 
and the corresponding rank two tensor integral {\it ggtt1} with $k_1\cdot k_2$ in the numerator. 
We vary $s_{12}$ and fix $s_{23}=-1.25, m_2=m_1, p_3^2=p_4^2=m_1^2=1$.
}
\label{fig:ggtt1}
 \end{figure}

Numerical results for the diagram {\it ggtt1} (see Fig.~\ref{fig:ggtt2}(a)) are shown in 
Fig.~\ref{fig:ggtt1} for both the scalar integral and an irreducible
rank two tensor integral. 

The integral representation of the diagram {\it ggtt1} is given by
\begin{align}
\label{eq:ggtt1int}
\mathcal{G}_{ggtt1}= &\left( \frac{1}{\mathrm{i} \pi^{\frac{\mathrm{D}}{2}}}\right)^2 
 \int \frac{\mathrm{d^D}k_1\, \mathrm{d^D}k_2}{D_1\ldots D_7}\;\,;\, D_1=k_1^2-m_2^2, D_2=(k_1+p_1)^2-m_2^2,\\
&  D_3=k_2^2-m_2^2, D_4=(k_2+p_2)^2-m_2^2, D_5=(k_1-k_2+p_1)^2, \nn\\
& D_6=(k_1-k_2-p_2)^2, D_7=(k_1-k_2+p_1+p_3)^2-m_1^2 \nn
  \;,
\end{align}
where we omitted the infinitesimal $\mathrm{i} \delta$ in the propagators, and we use the convention that 
all external momenta are ingoing. The momenta $p_3$ and $p_4$ are massive on-shell: $p_3^2=p_4^2=m_1^2$. 
The results shown Fig.~\ref{fig:ggtt1}(b) 
correspond to the rank two  tensor integral with the same propagators and 
a factor of $k_1\cdot k_2$ in the numerator. The numerical integration errors are 
shown as horizontal markers on the vertical lines. The absence of such markers means that 
the numerical errors are smaller than visible in the plot.

The timings for one kinematic point for the scalar integral in Fig.\,\ref{fig:ggtt1}(a) 
range from 11-60 secs for points far from threshold to $1.6\times 10^3$ 
seconds for a point very close to threshold, with an average of about 500 secs 
for points in the vicinity of the threshold.
A relative accuracy 
of $10^{-3}$ has been required for the numerical integration, 
while the absolute accuracy has been set to  $10^{-5}$. 
For the tensor integral, the timings are  better than in the scalar case, 
as the numerator function present in this case smoothes out the singularity structure. 
A phase space point far from threshold takes about 5-10 secs, while  points 
very close to threshold do not exceed 3600 secs
for the rank 2 tensor integral, Fig.\,\ref{fig:ggtt1}(b). 
The timings were obtained on a single machine using Intel i7 processors and 8 cores.

\begin{figure}[htb]
\begin{center}
\includegraphics[width=8.cm]{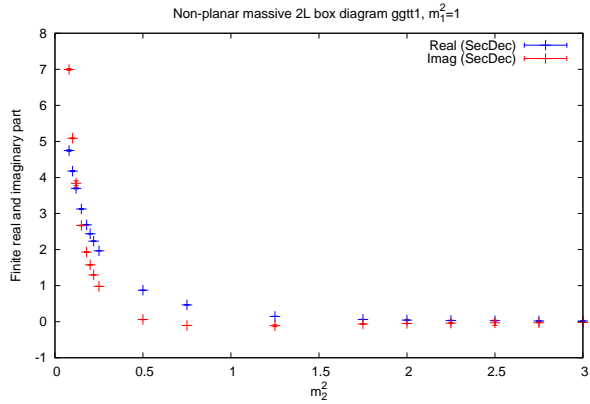} 
\caption{Results for the scalar integral  {\it ggtt1} shown in Fig.~\ref{fig:ggtt2}(a), 
with two different masses. We vary $m_{2}$ and fix $s_{12}=5, s_{23}=-1.25, p_3^2=p_4^2=m_1^2=1$.}
\label{fig:ggtt1m1m2}
\end{center}
 \end{figure}

For the results shown in  Fig.~\ref{fig:ggtt1} we used 
the numerical values $m_1^2=m_2^2=m^2=1, s_{23}=-1.25, s_{13}=2\,m^2-s_{12}-s_{23}$.
We set $m_1^2=m_2^2$ for the results shown in Fig.~\ref{fig:ggtt1} 
because this is the only case occurring in the process $gg\to t\bar{t}$ at two loops
if the $b$-quarks are assumed to be massless. 
Some numerical results for  kinematic points with $m_1\not=m_2$ are shown in Fig.~\ref{fig:ggtt1m1m2} in order to demonstrate that it is easy to add another 
mass scale in our approach, while this would complicate analytical calculations enormously.

\subsection{The {\it ggtt2} diagram}

\begin{figure}[htb]
\subfigure[leading pole]{	      
\includegraphics[width=0.5\textwidth]{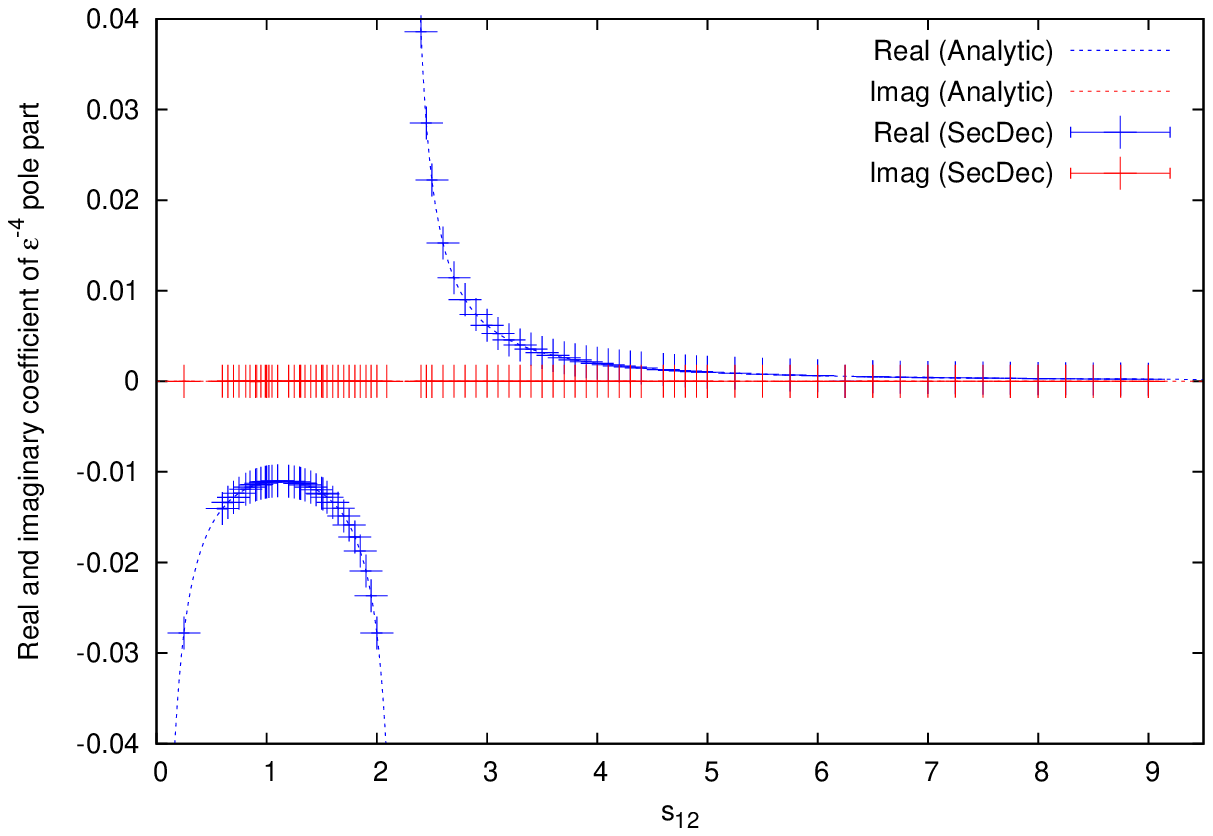}}\hfill
\subfigure[subleading pole]{
\includegraphics[width=0.5\textwidth]{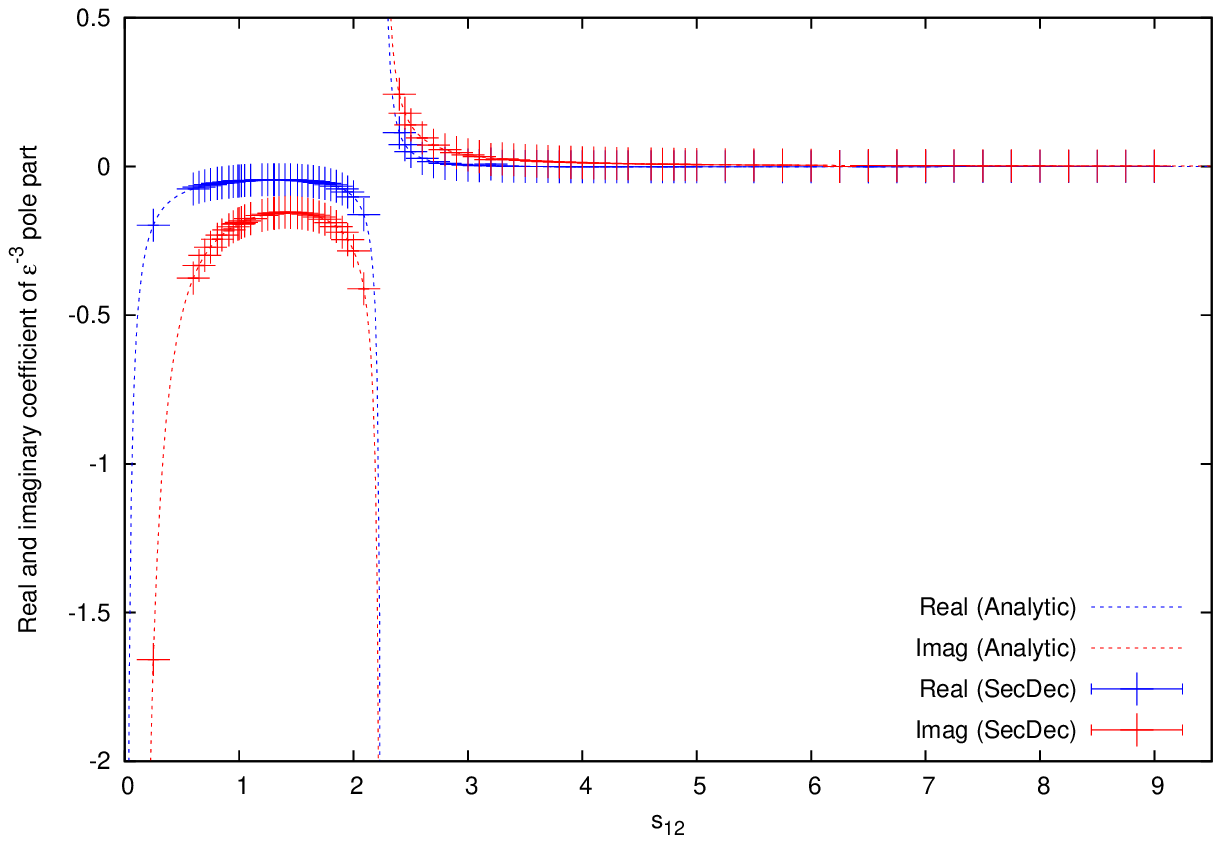}
}
	\caption{Comparison of (a) the leading and (b) the subleading pole coefficients between 
	the analytic result from \cite{AvMACAT} and the \secdec result, 
	for both real part (blue) and imaginary part (red).
We vary $s_{12}$ and fix $s_{23}=-1.25, p_3^2=p_4^2=m^2=1$.
	\label{fig:ggtt2_poles43}} 
\end{figure}

\begin{figure}[htb]
\subfigure[$1/\eps^2$ pole]{
	      \includegraphics[width=0.5\textwidth]{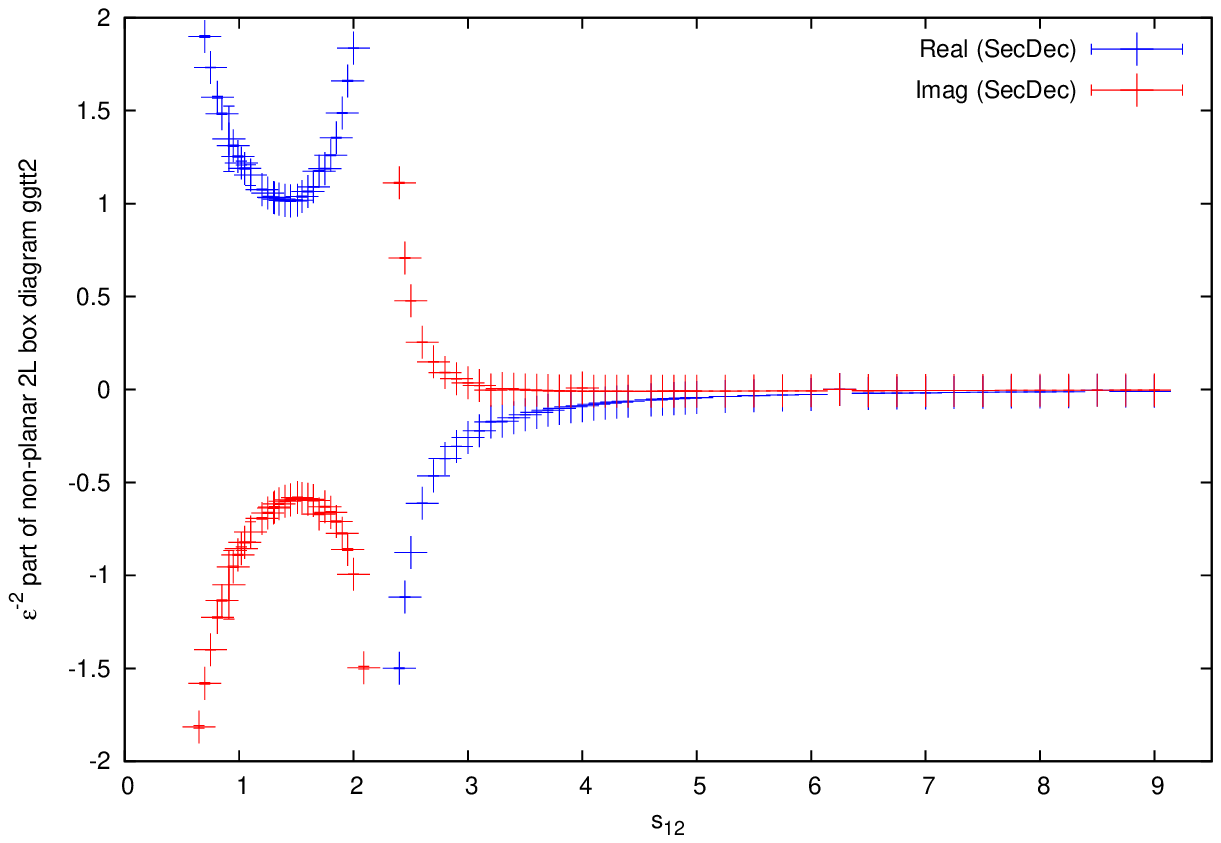}
}\hfill
\subfigure[$1/\eps$ pole]{	      
	      \includegraphics[width=0.5\textwidth]{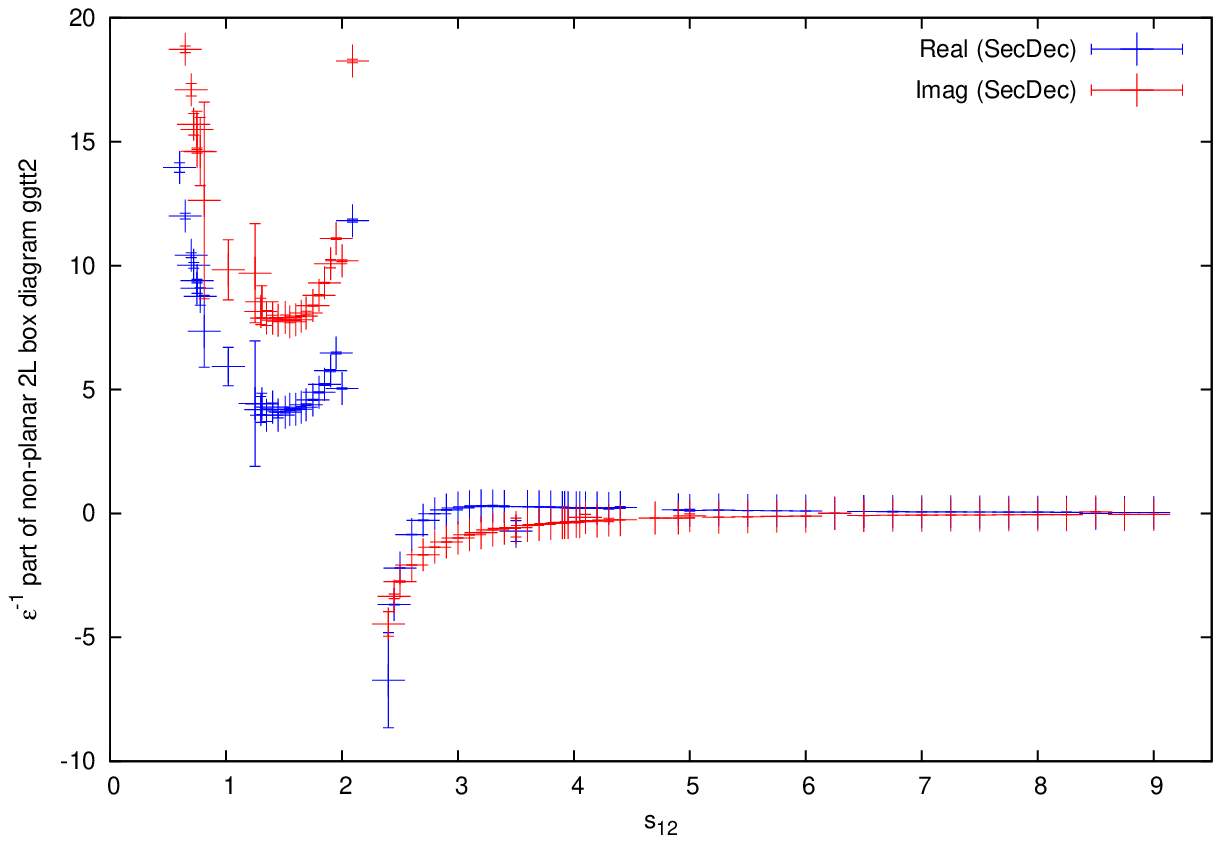}
}	      
\caption{Results for (a) the $1/\eps^2$ and (b) the $1/\eps$ coefficients of the integral  {\it ggtt2}.
The kinematics are the same as in Fig. \ref{fig:ggtt2_poles43}.}
\label{fig:ggtt2_poles21} 
\end{figure}

\begin{figure}[htb]
\subfigure[finite part, including threshold]{
\includegraphics[width=0.5\textwidth]{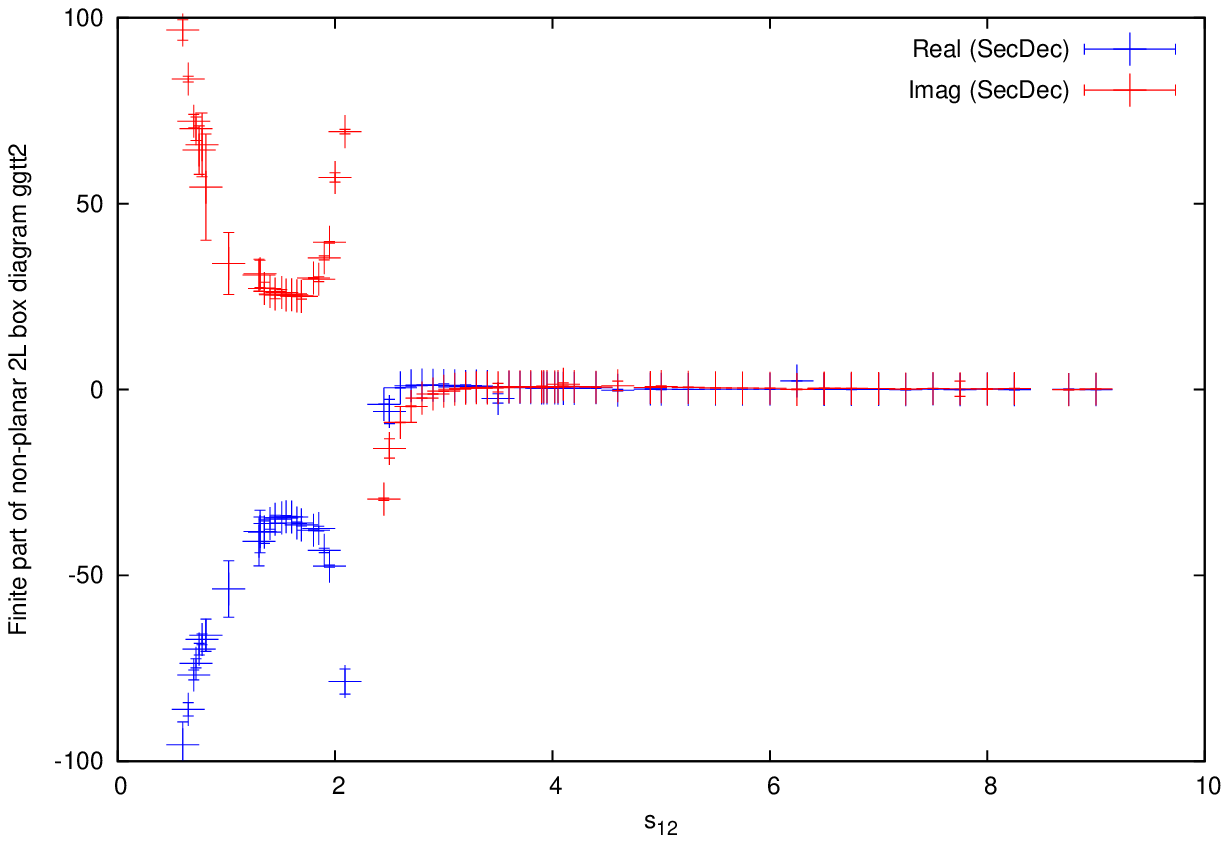} }
\hfill
\subfigure[finite part, beyond threshold]{
\includegraphics[width=0.5\textwidth]{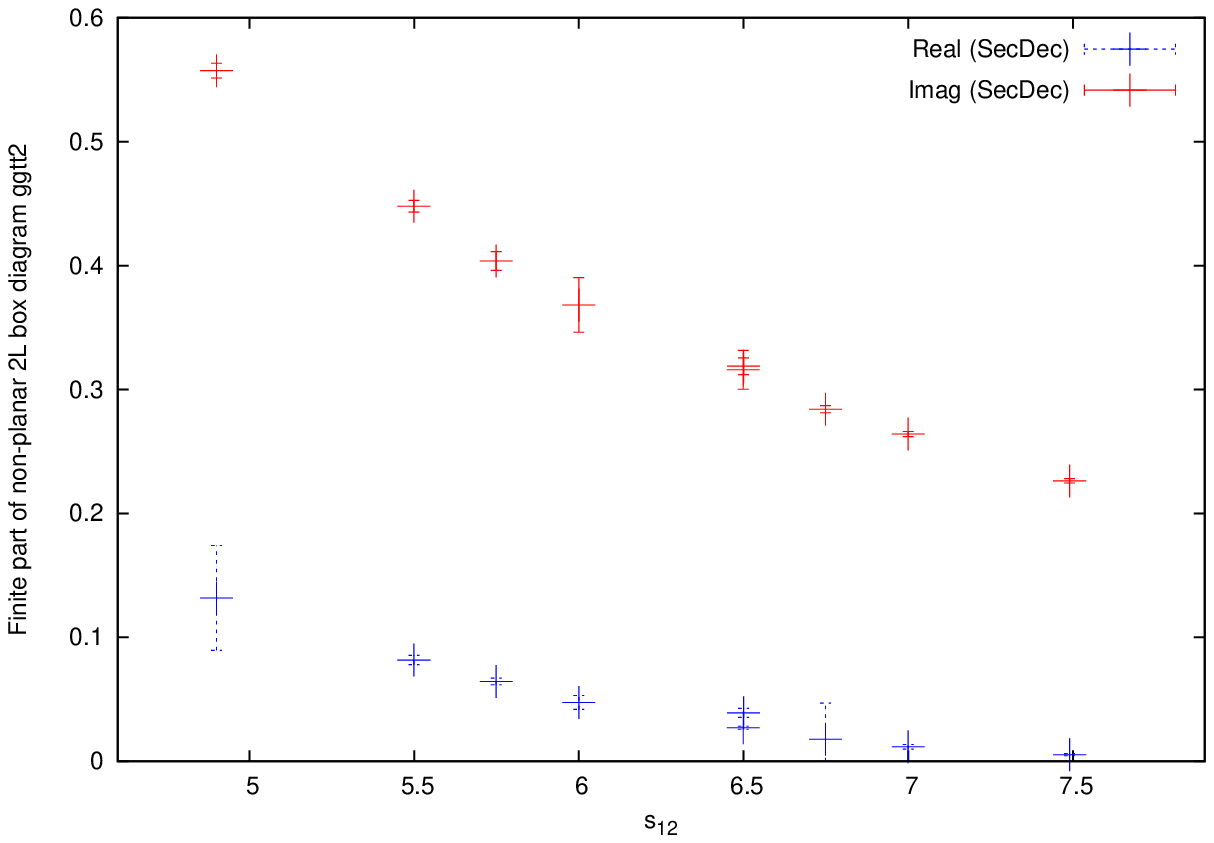} }
\caption{Results for the finite part of the scalar integral  {\it ggtt2}, 
(a) for a larger kinematic range, (b) zoom into a region further away from threshold.
The kinematics are the same as in Fig. \ref{fig:ggtt2_poles43}. 
The vertical bars denote the numerical integration errors.}
\label{fig:ggtt2finite}
\end{figure}

Analytic results for the pole coefficients of the $1/\eps^4$ and $1/\eps^3$ part 
of the diagram {\it ggtt2} shown in Fig.~\ref{fig:ggtt2}(b) have been given in \cite{AvMACAT}. 
We confirm these results and show a comparison between analytic and numerical results in 
Fig.~\ref{fig:ggtt2_poles43}.
The integral representation of the diagram {\it ggtt2} is given by eq.~(\ref{eq:ggtt1int}) with $m_2=0$. 
For the results shown in  Figs.~\ref{fig:ggtt2_poles43} to \ref{fig:ggtt2finite} we used 
the numerical values $p_3^2=p_4^2=m^2=1, s_{23}=-1.25, s_{13}=2m^2-s_{12}-s_{23}$, 
and we extract an overall factor of $-16\,\,\Gamma(1+\eps)^2$. 

For two selected phase space points, we also compared the full result for
all Laurent coefficients including  the finite part to the result obtained 
analytically~\cite{AvMprivate} and find agreement within the numerical integration errors.
Our numerical results for the remaining pole coefficients and the 
finite part of the diagram {\it ggtt2} are shown in 
Figs.~\ref{fig:ggtt2_poles21} and \ref{fig:ggtt2finite}. 

The timings for the leading and subleading pole coefficients of the diagram {\it ggtt2} are
ranging between fractions of a second and about 20 seconds. 
The coefficients of the $1/\eps^2$ pole take between 13 and 300 seconds, while the 
coefficients of the $1/\eps$ pole take between 75 and 3000 seconds, depending 
on their distance to thresholds. 
For the finite part, the integration times range from 250 seconds to 4000 seconds.
For all Laurent coefficients, a relative accuracy of $5\times 10^{-3}$ has been required,
which has not always been reached  for the $1/\eps$ and $\eps^0$ coefficients.
It also should be noted that the timings for points close to threshold depend rather sensitively on the 
settings for the Monte Carlo integration parameters.

\subsection{Massive tensor two-loop two-point functions}

\begin{figure}[htb]
\begin{center}
\includegraphics[width=5.2cm]{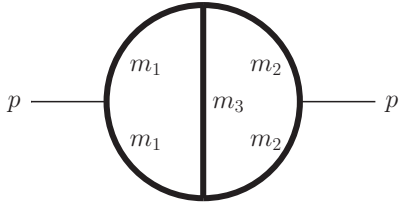} 
\caption{Two-loop bubble diagram with  different masses}
\label{fig:bubble2m}
\end{center}
\end{figure}

Here we show that the option to evaluate integrals with a non-trivial numerator 
can also be applied to calculate two-loop two-point functions involving different 
mass scales, without the need 
for a reduction to master integrals. This fact can be used for instance to calculate 
two-loop corrections to mass parameters in a straightforward way.
As an example we pick the diagram shown in Fig.~\ref{fig:bubble2m}.

\begin{figure}[htb]
\subfigure[scalar integral]
{{\includegraphics[width=0.53\textwidth]{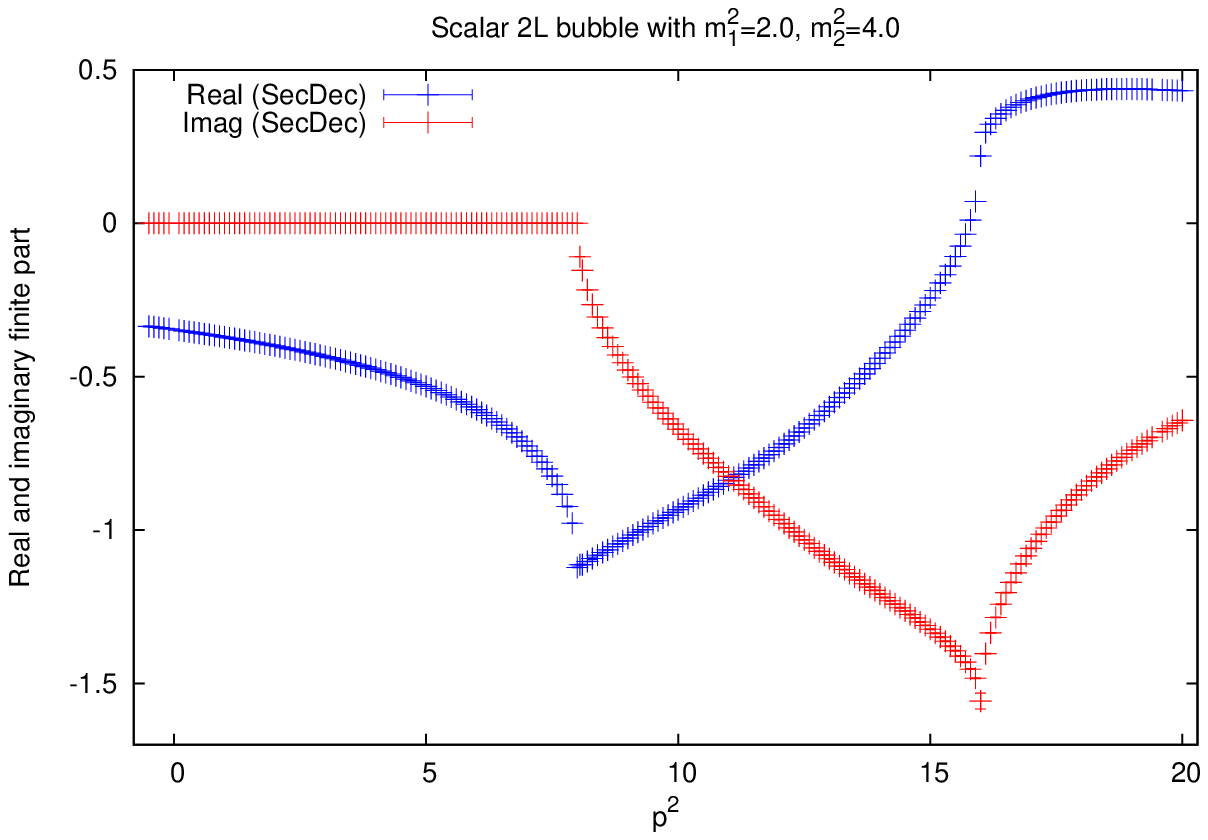}} }\hfill
\subfigure[rank 3 tensor integral]
{\includegraphics[width=0.53\textwidth]{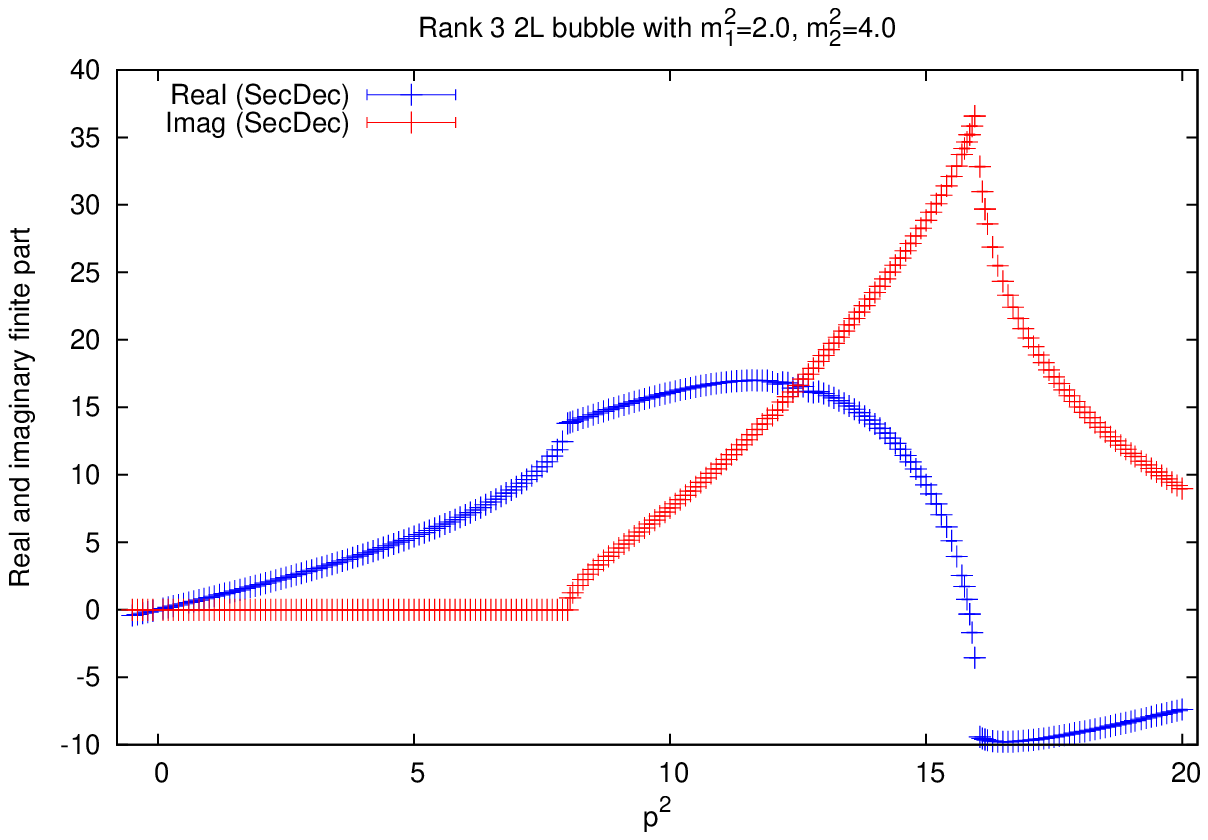} }
\caption{Results for real (blue) and imaginary (red) parts of 
the rank 3 two-loop bubble diagram shown in Fig.\,\ref{fig:bubble2m}, 
(a) scalar case, (b) with numerator $(k_1\cdot k_2)\,(k_1\cdot p_1)$. 
The masses are $m_1^2=2, m_2^2=4, m_3=0$.}
\label{fig:bub_m3zero}
\end{figure}

\begin{figure}[htb]
\subfigure[scalar integral]
{\includegraphics[width=0.5\textwidth]{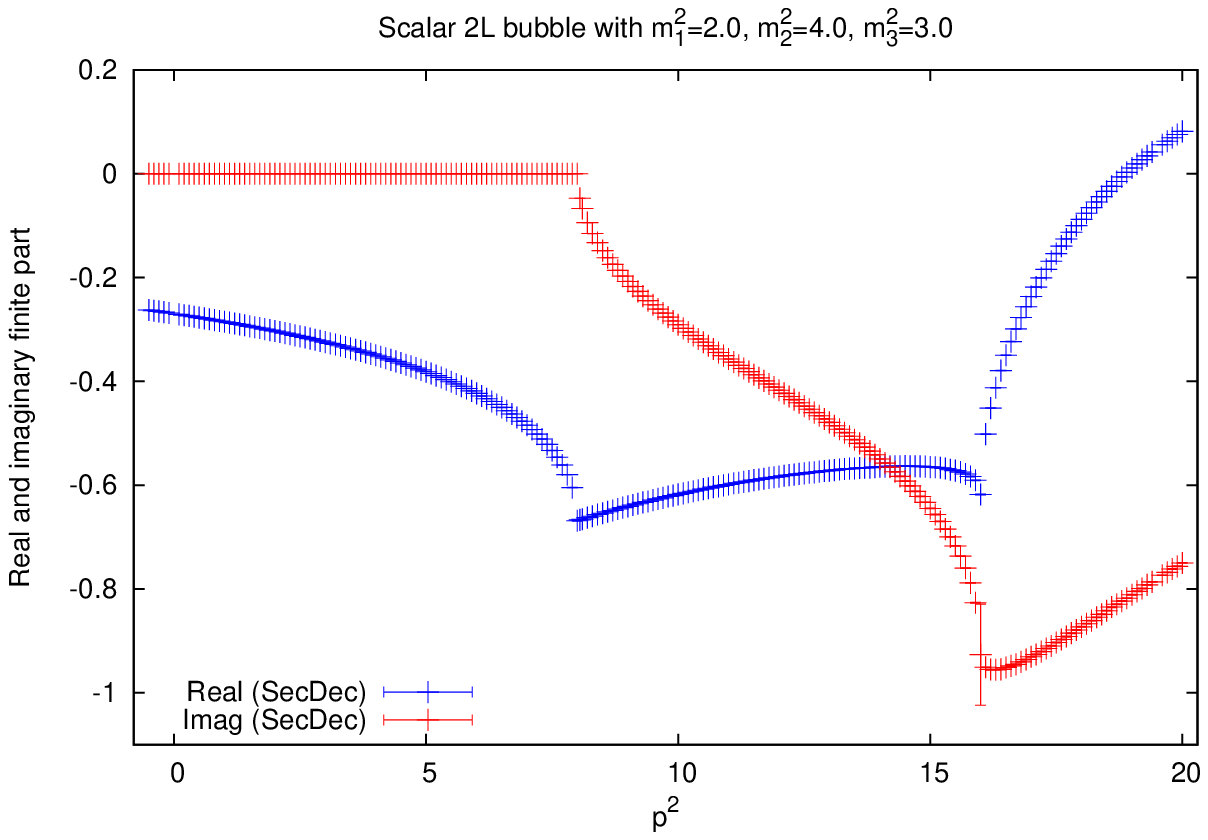}} \hfill
\subfigure[rank 3 tensor integral]
{\includegraphics[width=0.5\textwidth]{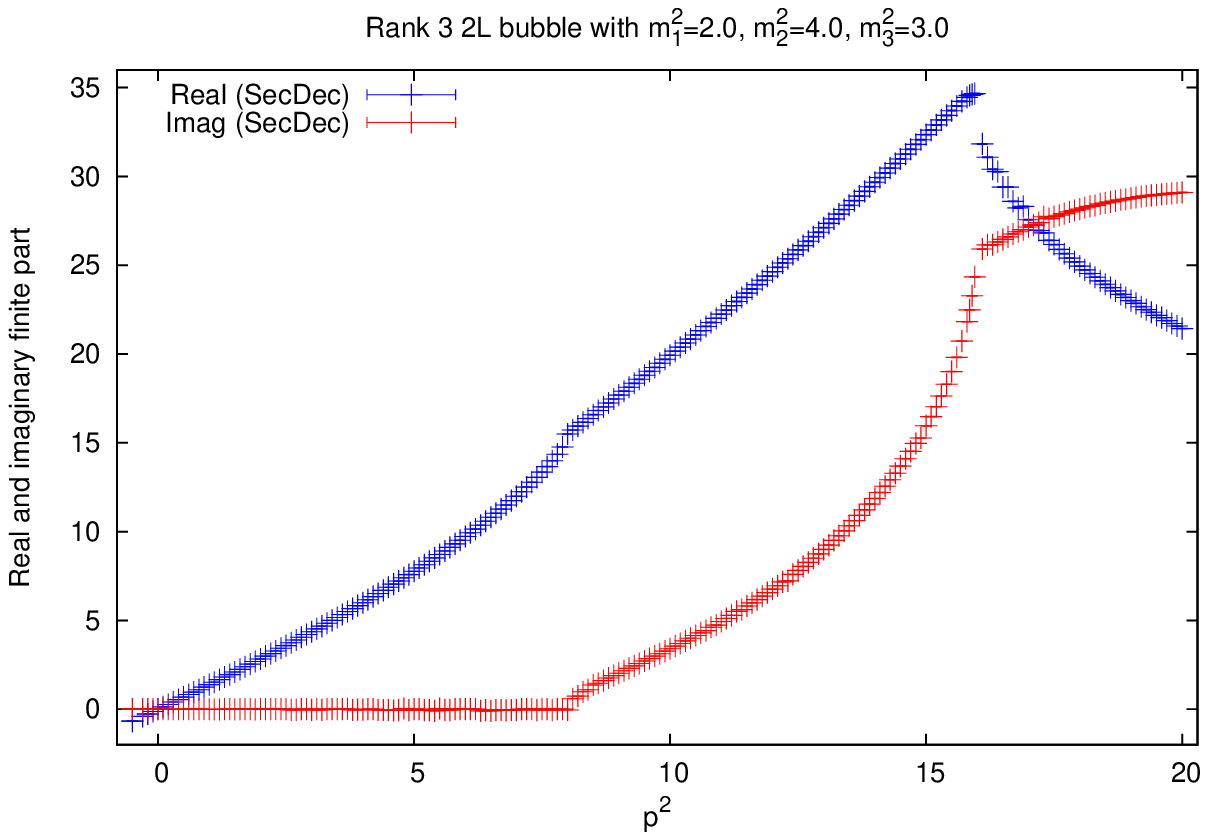} }
\caption{Results for the rank 3 two-loop bubble diagram with three non-vanishing masses,
shown in Fig.\,\ref{fig:bubble2m}, 
(a) scalar case, (b) tensor case with numerator $(k_1\cdot k_2)\,(k_1\cdot p_1)$. 
The masses are $m_1^2=2, m_2^2=4, m_3^2=3$.} 
\label{fig:bubble3m} 
\end{figure}

We calculate scalar integrals and  rank three tensor integrals, for two cases, 
one where $m_3$ is zero, and one where  $m_3$ is nonzero.
The tensor integral is given by
\begin{align}
\mathcal{G}_{B} &= \left( \frac{1}{\mathrm{i} \pi^{\frac{\mathrm{D}}{2}}}\right)^2 
 \int \frac{\mathrm{d^D}k_1\, \mathrm{d^D}k_2\,\, (k_1\cdot k_2)\,(k_1\cdot p_1)}{D_1\ldots D_5}\;\,; \,
\label{eq:bubbleint}\\
& D_1=k_1^2-m_1^2, D_2=(k_1+p_1)^2-m_1^2, D_3=(k_1-k_2)^2-m_3^2, \nn\\
& D_4=(k_2+p_1)^2-m_2^2, D_5=k_2^2-m_2^2 \;.\nn
\end{align}
The fact that this tensor integral is reducible does not play a role here, because our purpose is
to demonstrate that reduction may become obsolete considering the short integration times for the tensors.
Results for the scalar and tensor integrals with $m_3=0$ are shown in Fig.~\ref{fig:bub_m3zero}, 
while results for $m_3^2=3$ are shown in Fig.~\ref{fig:bubble3m}.

\begin{figure}[htb]
\subfigure[$m_3=0$]{
\includegraphics[width=0.5\textwidth]{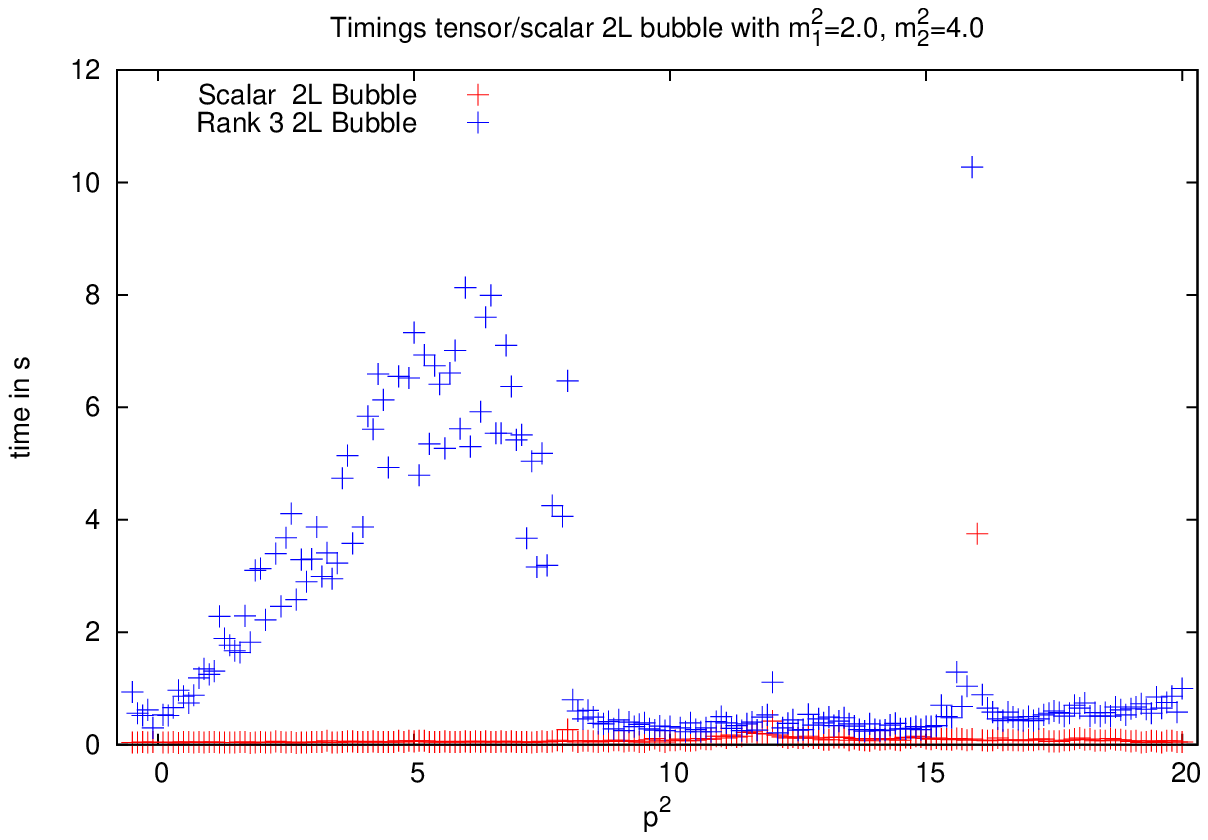} }
\subfigure[$m_3^2=3$]{
\includegraphics[width=0.5\textwidth]{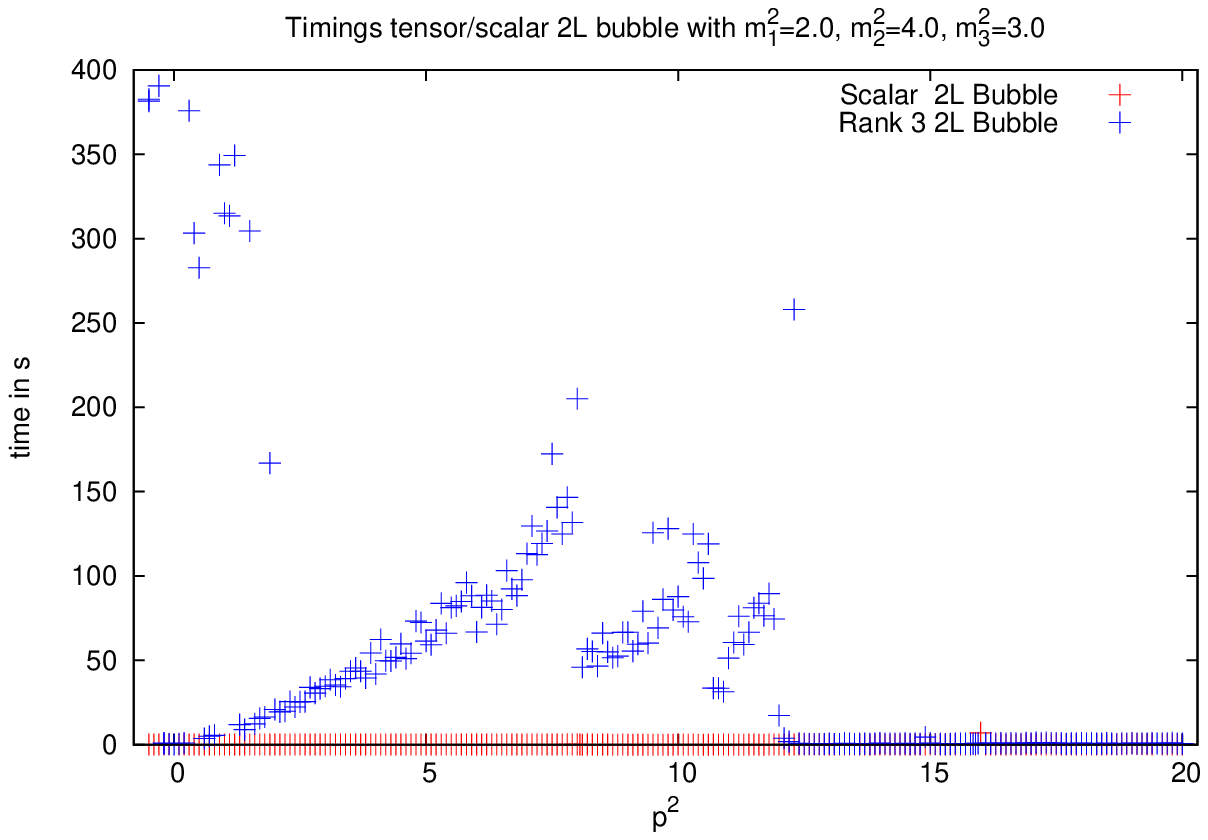} 
}
\caption{Comparison of evaluation times between the scalar and the rank three tensor integrals
corresponding to a two-loop two-point function with 3 different masses.
The red points are the evaluation times in seconds for the scalar integral at a given 
kinematic point, the timings for the rank 3 tensor integral are marked in blue.
\label{fig:bubble3m_timings}
}
\end{figure}

The timings for the massive two-point integrals  
are shown  in Figs.~\ref{fig:bubble3m_timings}\,(a) and (b), 
for  $m_3^2=0$ and $m_3^2=3$, respectively.
The timings were obtained on computers with Intel i7 processors and 8 cores.
In both cases, a relative accuracy of 0.1\% was required for the 
Monte Carlo integration. In Fig.\,\ref{fig:bubble3m_timings}\,(a),
an absolute
desired accuracy of $10^{-3}$ was set, 
in Fig.\,\ref{fig:bubble3m_timings}\,(b) an absolute accuracy of
$10^{-6}$.
Fig.\,\ref{fig:bubble3m_timings} shows that for values of $p^2$ 
above the mass threshold at $p^2=4m_2^2$, 
the timings for the contracted rank three 
tensor integrals do not differ much from the ones for the scalar integrals.
In the region below threshold, the timings  
are higher because  the imaginary part is zero, and a vanishing 
function is difficult to integrate numerically. 
As the relative error to a zero value is always infinite, the numerical
integrator in this case tries to reach the absolute accuracy goal.
If in addition to the vanishing imaginary part the real part is also 
close to zero, the integration times are highest, as can be seen from 
Figs.\,\ref{fig:bubble3m} and \ref{fig:bubble3m_timings}\,(b).
To avoid artificially large timings in kinematic regions where 
the imaginary part is known to be zero, one could set {\tt contourdef=false}
for the kinematic points below threshold. This would have the effect that the 
imaginary part is not calculated at all, but set to zero from scratch, 
thereby reducing the integration time considerably.
The minimal numerical integration time for a kinematic point above threshold in the case
of the scalar two-loop bubble integrals is 0.03\,secs for $m_3^2\not=0$ and 0.02\,secs for
$m_3^2=0$.

\vspace*{5mm}

\section{Conclusions}
\label{sec:conclusion}
We have presented numerical results for two massive non-planar seven propagator topologies, 
one entering  the light fermionic two-loop correction to the $gg\to t\bar{t}$ channel,  
the other one entering  the heavy fermionic correction to this channel. 
For the latter, no analytical result is available yet. 
Apart from the scalar master integral, we also give results for an irreducible 
tensor numerator of rank two for this diagram.
Our numerical results have been obtained with the program {\sc SecDec}2.1, 
which is publicly available at\\ {\tt http://secdec.hepforge.org}.
Compared to version 2.0 of {\sc SecDec}, version 2.1 contains a number of new
features, among them the possibility to evaluate 
tensor integrals with  no principle limitation on the rank. 
The applicability of the tensor option to various types of integrals is further 
demonstrated by a number of results for two-loop two-point functions involving several different mass scales, 
where no analytical results exist.
Another new feature is the option to apply the sector decomposition algorithm and subsequent contour 
deformation on user-defined functions which do not necessarily have the form of standard loop integrals. 
This new feature is used in combination with a novel type of analytic transformations, which can serve to 
reduce the number of functions to be integrated numerically. 
We believe that \secdec version 2.1 brings us a major step forward 
in moving from the calculation of master integrals 
to the calculation of two-loop corrections for phenomenological applications.

\section*{Acknowledgements}
We would like to thank Andreas von Manteuffel for sending us his results for 
the non-planar box diagram ggtt2 and  for very useful discussions.
We are also grateful to Michael Gustafsson for very valuable comments about the usage of the program.
\renewcommand \thesection{\Alph{section}}
\renewcommand{\theequation}{\Alph{section}.\arabic{equation}}
\setcounter{section}{0}
\setcounter{equation}{0}
\section{Appendix}
\subsection{Usage of the program}
\begin{enumerate}
\item Change to the subdirectory {\tt loop} or {\tt general}.
The setup in the {\tt loop} directory should be used 
to calculate loop integrals and integrals with a similar structure. 
The option to use contour deformation is available for all functions processed within the 
{\tt loop} directory.
The setup in the {\tt general} directory allows to evaluate more general parameter integrals, 
which can have endpoint singularities at zero and one, 
but does not offer contour deformation.
\item In the {\tt general} directory, copy the files {\tt param.input} and \\
{\tt template.m} to 
create your own parameter and template files \\ {\tt myparamfile.input}, {\tt mytemplatefile.m}
and edit them according to the function to be calculated.
In the {\tt loop} 
directory, edit the files \\
{\tt paramloop.input} and {\tt templateloop.m} if you want to compute a Feynman loop
integral in a fully automated way.
If you would like to define a set of own functions rather than a standard loop integral, 
use the files {\tt paramuserdefined.input} and {\tt templateuserdefined.m} as a starting point.
\item Set the desired parameters in {\tt myparamfile.input} and specify the integrand in {\tt mytemplatefile.m}.
\item Execute the command {\it ./launch -p myparamfile.input -t mytemplatefile.m} 
in the shell. If you add the option {\it -u } in the {\tt loop} directory, user defined functions are computed. \\
If you omit the option {\it -p myparamfile.input}, the file {\tt param.input} will be taken as default.
Likewise, if you omit the option {\it -t mytemplatefile.m}, 
the file {\tt template.m} will be taken as default.
If your files {\it myparamfile.input, mytemplatefile.m} are in a different directory, say, 
{\it myworkingdir}, 
 use the option {\bf -d myworkingdir}, i.e. the full command then looks like 
 {\it ./launch -d myworkingdir -p myparamfile.input -t mytemplatefile.m}, 
 executed from the directory {\tt SecDec/loop} or
 {\tt SecDec/general}.
 \item Collect the results. Depending on whether you have used a single machine or 
submitted the jobs to a cluster, the following actions will be performed:
 \begin{itemize}
\item If the calculations are done sequentially on a single machine, 
    the results will be collected automatically (via the corresponding {\tt results*.pl} called by {\tt launch}).
    The output file will be displayed with your specified text editor.
\item If the jobs have been submitted to a cluster,    
	when all jobs have finished, execute the command 
	{\it ./results.pl [-d myworkingdir -p myparamfile]} in the general, and 
	{\it ./resultsloop.pl [-d myworkingdir -p myparamfile]} or
	{\it ./resultsuserdefined.pl [-d myworkingdir -p myparamfile]} in the loop directory, respectively.
	This will create the files containing the final results in the {\tt graph} subdirectory
	specified in the input file.
\end{itemize}
\item After the calculation and the collection of the results is completed, 
you can use the shell command {\it ./launchclean[graph]}
to remove obsolete files.
\end{enumerate}
\subsection{Evaluation of user-defined functions in the loop directory}
\label{appendix:userdefined}
In the following, we will describe the input and syntax needed for the files
\\{\tt mytemplatefile.m} and  {\tt myparamfile.input} 
when treating functions which are different from the standard Feynman parameter representation 
which -- in the default setup -- is derived automatically from 
the propagators or vertices of a multi-loop integral. 

The file {\tt mytemplatefile.m} should contain the following information:
\begin{itemize}
  \item {\bf List of user-defined functions:}\\
The user-defined functions should be polynomial in the Feynman parameters. 
They can contain monomial 
factors of Feynman parameters with arbitrary exponents, 
functions of type  $\mathcal{U}$ and  $\mathcal{F}$ (i.e. polynomials in the Feynman parameters
involving also kinematic invariants) with arbitrary exponents 
and a ``numerator" with positive exponents only. 
An iterated sector decomposition is applied to $\mathcal{U}$ and
    $\mathcal{F}$   if the decomposition flag (see below)
    is set to {\tt B}. In case the functions $\mathcal{U}$ and $\mathcal{F}$ contain thresholds, 
    a deformation of the integration contour into the complex plane becomes necessary. 
The setup is such that
the integration contour will be formed based on the function $\mathcal{F}$.
 The list of user-defined functions must be inserted into  {\tt mytemplatefile.m} using the following syntax\\
\texttt{functionlist}=\{{\it function\_1},{\it function\_2},...,{\it function\_i},...\};\\
  with \\
{\it function\_i}=\{{\it \# of function},\{{\it list of exponents}\},\{\{{\it function $\mathcal{U}$}, {\it exponent of $\mathcal{U}$}, {\it decomposition flag}\},\{{\it function $\mathcal{F}$},{\it exponent of $\mathcal{F}$},{\it decomposition flag}\}\}, {\it numerator}\}.\\
The {\it \# of function} of each user-defined function is a label of the function by an integer, 
where the default is just sequential numbering.
However, there is also the option to label a set of functions with the same integer. In this case 
the functions are decomposed individually, but will be combined  after the decomposition.
This leads to fewer or simpler integrand functions when symmetries are found within a sector. 
It should be noted however that only functions which share the same exponent for all functions of type
$\mathcal{U}$ respectively $\mathcal{F}$ can be grouped together and therefore 
have the same function label.

Each entry in the comma separated {\it list of exponents} corresponds to an exponent of a 
Feynman parameter occurring as a monomial in the Feynman integral.
The {\it decomposition flag} should be  {\tt A} if no iterated sector decomposition is desired, 
and  {\tt B} if the function needs further decomposition. 
The {\it numerator} may contain several functions,  with different non-negative exponents.
\item {\bf Dimension, kinematic conditions:}\\
The space-time dimension should be specified in {\tt mytemplatefile.m} via \\
{\tt Dim=}{\it dimension}.
The default is {\tt Dim=}$4-2\eps$.\\
On-shell conditions can be specified in the list {\tt onshell=\{\}}.
  %
\item {\bf Computation of the exponents of $\mathcal{F}$ and $\mathcal{U}$ (optional):}\\
If the exponents of the functions $\mathcal{F}$ and $\mathcal{U}$ should be computed by the program
according to the rules for standard multi-loop integrals, the user needs to specify
the number of propagators and their powers in a list {\tt powerlist=Table[}{\it power}{\tt,\{i,}{\it \#propagators}{\tt \}];} 
and the tensor rank of the diagram via {\tt rank=}{\it rank}{\tt ;}\hspace{5pt}.\\
Here, {\it \#propagators} should correspond to the number of propagators of the original Feynman diagram, 
or, more general, to the number of integration variables plus one.
 \end{itemize}
In addition to the changes made to  {\tt mytemplatefile.m}, it is possible to set a maximal number of Feynman parameters occurring in the user defined functions by initalizing
{\tt feynpars=\ldots} in  {\tt myparamfile.input}. The default is the number of propagators $N$ subtracted by one. \\ 
A full example including detailed comments comes with the download of the program and can be found in {\tt /loop/templateuserdefined.m} and \\{\tt /loop/paramuserdefined.input}.
\subsection{Evaluation of tensor integrals}
\label{appendix:tensor}
For the computation of tensor integrals, where the tensor is contracted with external momenta and/or loop momenta, 
the construction of the Feynman integral via topological cuts needs to be switched off, 
which corresponds to {\tt cutconstruct=0}. \\
Regarding the user input, the only additional information needed is the numerator in  {\tt mytemplatefile.m}. 
Each scalar product of loop momenta contracted with either external momenta or 
other loop momenta should be given as an entry of a list:\\
{\tt numerator=\{{\it prefactor},{\it comma separated list of scalar products}\}}.
For example, a numerator of the form  -$2\,k_1\cdot k_2$,  where $k_1$ and $k_2$ are loop momenta, 
should be given  as 
{\tt numerator=\{}$-2$, $k1*k2$ {\tt \}}.
A numerator of the form $2\,(k_1\cdot p_1)(k_1\cdot k_2)$, where $p_1$ is an external momentum, 
should be given as \\
{\tt numerator=\{}2, $k1*p1$, $k1*k2$ {\tt \}}.

\label{sec:appendix}
 

\providecommand{\href}[2]{#2}\begingroup\raggedright\endgroup

\end{document}